\documentclass[final,authoryear,10pt,twocolumn]{elsarticle}

\usepackage{graphicx}

\usepackage{amssymb}

\journal{Journal}

\begin{document}

\begin{frontmatter}

\title{A Flexible System for Automatic Quality Control of Oceanographic Data}

\author[iousp,scripps]{Guilherme P. Castel\~{a}o\corref{gui}}
\ead{guilherme@castelao.net}
\ead[url]{http://cotede.castelao.net}
\cortext[gui]{Corresponding author}

\address[iousp]{Instituto Oceanogr\'{a}fico -- USP, S\~{a}o Paulo, Brazil.}
\address[scripps]{Now at Scripps Institution of Oceanography -- UC San Diego, California, USA}

\begin{abstract}

Sampling errors are inevitable when measuring the ocean;
thus, to achieve a trustable set of observations requires a quality control (QC) procedure capable to detect spurious data. 
While manual QC by human experts minimizes errors, it is inefficient to handle large datasets and vulnerable to inconsistencies between different experts. 
Although automatic QC circumvents those issues, the traditional methods results in high rates of false positives. 
Here, I propose a novel approach to automatically QC oceanographic data based on the anomaly detection technique. 
Multiple tests are combined into a single, multidimensional criterion that learns the behavior of the good measurements, and identifies bad samples as outliers. 
When applied to 13 years of hydrographic profiles, the anomaly detection resulted in the best classification performance, reducing the error by at least 50\%. 
An open source Python package, CoTeDe, was developed to provide state of the art tools to quality control oceanographic data.
\end{abstract}

\begin{keyword}

quality control \sep temperature \sep salinity \sep CTD \sep ARGO \sep XBT \sep TSG \sep PIRATA \sep anomaly detection \sep fuzzy logic \sep machine learning

\end{keyword}

\end{frontmatter}


\section{Introduction}\label{sec:introduction}

Conservation of momentum, heat, and mass in the ocean depend on the seawater density ($\rho$) (see \citet{Gill1982}), thus $\rho$ is a necessary variable to describe and to understand processes at the most broad range of scales: from global sea level rise to oil and pollutant dispersion. 
Because variations of $\rho$ are usually small, together with relatively large accelerations in the ocean, it becomes impractical for an instrument to make direct {\it in situ} density measurements along the water column \citep{Baker1981}. 
As an alternative, oceanographers infer $\rho$ from temperature ($T$), salinity ($S$), and pressure ($P$), using a Gibbs function formulation for seawater \citep{Feistel2008, Millero2008, Baker1981}.
The \{T, S, P\} structure is a fundamental building block of physical oceanography, thus errors in describing those variables compromise any outcome conclusions: 
A persistent bias on $T$ profiles from eXpendable Bathy-Thermographs (XBT) resulted in up to 50\% error on the estimates of ocean warming and thermal expansion in recent decades \citep{Cowley2013, Domingues2008, Levitus2009}; 
Profiles from Argo floats lacking correction for pressure sensor drift misled to spurious variability on the \{$T$, $S$\} vertical structure \citep{Barker2011}.  
The robustness of a scientific study is tied to the quality of the data that grounded it.

The marine environment is harsh for electronic sensors, making inevitable to have some spurious measurements. 
In response to that, data distribution centers and coordinated observing programs have established clear quality control (QC) procedures, providing measurements with quality flags.  
Some of the widely used procedures were defined by: The Global Temperature and Salinity Profile Programme (GTSPP) \citep{gtsppqc2010}, the European Global Ocean Observing System (EuroGOOS) of the Intergovernmental Oceanographic Commission of UNESCO (IOC GOOS) \citep{EGOOS2010}, the Argo Program \citep{ARGO2015}, the CSIRO XBT Program \citep{Bailey1994}, the Australian Integrated Marine Observing System, and the Integrated Ocean Observing System \citep{QARTOD2016}.
While each type of sensor requires some specific steps in the QC procedure, the recommendations mentioned above have several tests in common and share the same general structure of a sequence of independent tests. 
A major weakness of this approach is the lack of context awareness to qualify the data \citep{Smith2012}, often compensated by complementary manual QC. 
Manual evaluation by experts is indeed the best current option to minimize errors, mostly due to the relatively limited amount of measurements in the oceans and the adaptive skills of the human brain in pattern recognition.
However, the efficiency of an automatic QC better suits realtime operations, such as weather forecast, as well as the processing of large datasets. 
Also, inconsistencies in manual QC can pose a problem to compare, or to integrate, multiples datasets \citep{Morello2014}. 
Thus, improvements in measuring the ocean must be followed by advances in oceanographic QC methods to reduce the burden on the human experts without compromising the classification skill, and to allow for efficient and coherent data aggregation.

Developments in data mining and machine learning have been revolutionizing data analysis (see \cite{IvezicConnollyVanderPlas2014} for a nice introduction on this subject). 
Such modern techniques provide a convenient framework to improve automatic QC by using supervised learning, which reduces the discrepancy with the human expert evaluation. 
From a machine learning perspective, oceanographic data QC is a classification task where the simplest setup is composed by only two categories: good or bad data. 
The most powerful technique used so far is Bayesian Networks \citep{Smith2012}, 
that combines different factors through an empirical network of relations to perform a quality classification by statistical inference. 
Such intricate decision making comes with a price: a slow learning rate that must outcome the natural variability. 
Therefore, to calibrate Bayesian Networks requires a larger volume of data and/or a higher sampling rate
\citep{Bettencourt2007}.
Coastal monitoring sensors with fixed positions may satisfy that requirement, allowing for a skillful QC classification \citep{Smith2012, Rahman2013, RahmanSmithTimms2014}, so that future improvements for similar scenarios shall come from tuning the Bayesian Network rather than employing a completely different technique.
That is not necessarily true for other types of sensors with different sampling strategies. 

Because a proper sampling procedure should minimize spurious measurements, the QC of hydrographic data is an unbalanced classification problem. 
The number of good measurements is typically at least two orders of magnitude larger than the available bad samples, which critically compromises the calibration of most machine learning techniques, including Bayesian Networks, Support Vector Machines, and Neural Networks.
\cite{RahmanSmithTimms2014} brought an important contribution by proposing to use cluster undersampling to circumvent the unbalancing issue, however that might not fully addresses the problem.
The data used on the calibration phase still must statistically represent each one of the categories being classified, but cluster undersampling cannot guarantee that. 
Spurious data have no bounds of feasibility, being, therefore, relatively more variable than valid data, yet representing the smallest fraction of the measurements.
If the available sample of bad data does not statistically represent all possible spurious measurements, the calibration would minimize the classification error for that particular dataset, but it would not necessarily be a good predictor for new bad samples, an issue known as overfitting \citep{IvezicConnollyVanderPlas2014}.
Even though cluster undersampling tackles the unbalancing issue, to calibrate a Bayesian Network still requires a sufficient amount of bad samples, which is unusual for oceanographic profiles which are sparse in space and time.

A machine learning technique more adequate to QC oceanographic data is the anomaly detection, which learns the behavior of the good data and identifies the bad data as an outlier. 
Therefore, neither the relative, nor the absolute, sample size of the bad data compromises the QC classification, allowing to identify spurious measurements even if that kind of error has never been observed before.
Previous implementations of anomaly detection for environmental measurements were based on the comparison of multiple sensors \citep{Bettencourt2007} or auto--regressive models \citep{HillMinsker2010}, which would not be the most appropriate for oceanographic profiles. 
In a novel approach, I propose to search for outliers on the characteristics of the data instead of the measured value itself. 
This is somehow aligned with the vision of \citet{Gronell2008}, but with the advantages of the anomaly detection framework.

Although the observing programs provide data with quality flags, all observational scientific studies require to quality control its own freshly collected data for its own use, even before providing it to the data centers. 
Non--operational studies can rarely afford a QC specialist in their team, creating an overhead and risking the quality of their data products and scientific conclusions. 
This manuscript introduces an open source Python package, named CoTeDe, that provides in a single tool the different possibilities on the state of the art to quality control oceanographic data. 
CoTeDe is optimized to attend data centers with large volumes of data, while flexible enough to accommodate a diverse combination of procedures and fine tuning required by specific studies. 
Such flexibility allows to provide the traditional QC approach, as well as modern powerful solutions like anomaly detection and fuzzy logic.
The performance of the different techniques are compared in a case study of CoTeDe on a real hydrographic dataset, described in Section \ref{sec:Data}. 
The traditional QC procedures are briefly reviewed in Section \ref{sec:Traditional}, the anomaly detection is introduced on Section \ref{sec:AnomalyDetection}, and a fuzzy logic approach based on \citet{Morello2014} is presented on Section \ref{sec:FuzzyLogic}.
The technical details on running CoTeDe are left for the user manual\footnote{http://cotede.castelao.net}.

\section{Methodology}

All techniques and QC tests discussed on this study were implemented as a collection of independent modules in the open source package CoTeDe, 
what makes it easier to extend for new tests, and gives the user flexibility to apply them. 
The user can customize any desired set of tests, including the specific parameters and thresholds of each test. 
Otherwise, there are preset QC procedures conforming with the recommendations from GTSPP \citep{gtsppqc2010}, EuroGOOS \citep{EGOOS2010} or ARGO \citep{ARGO2014}. 
In addition to that, it is also implemented innovating approaches based on Fuzzy Logic \citep{Timms2011, Morello2014} and Anomaly Detection. 
The different approaches to QC using CoTeDe are illustrated using the dataset described below. 

\subsection{Data}\label{sec:Data}

\begin{figure}[htb]
    \centering
    \noindent\includegraphics[width=20pc]{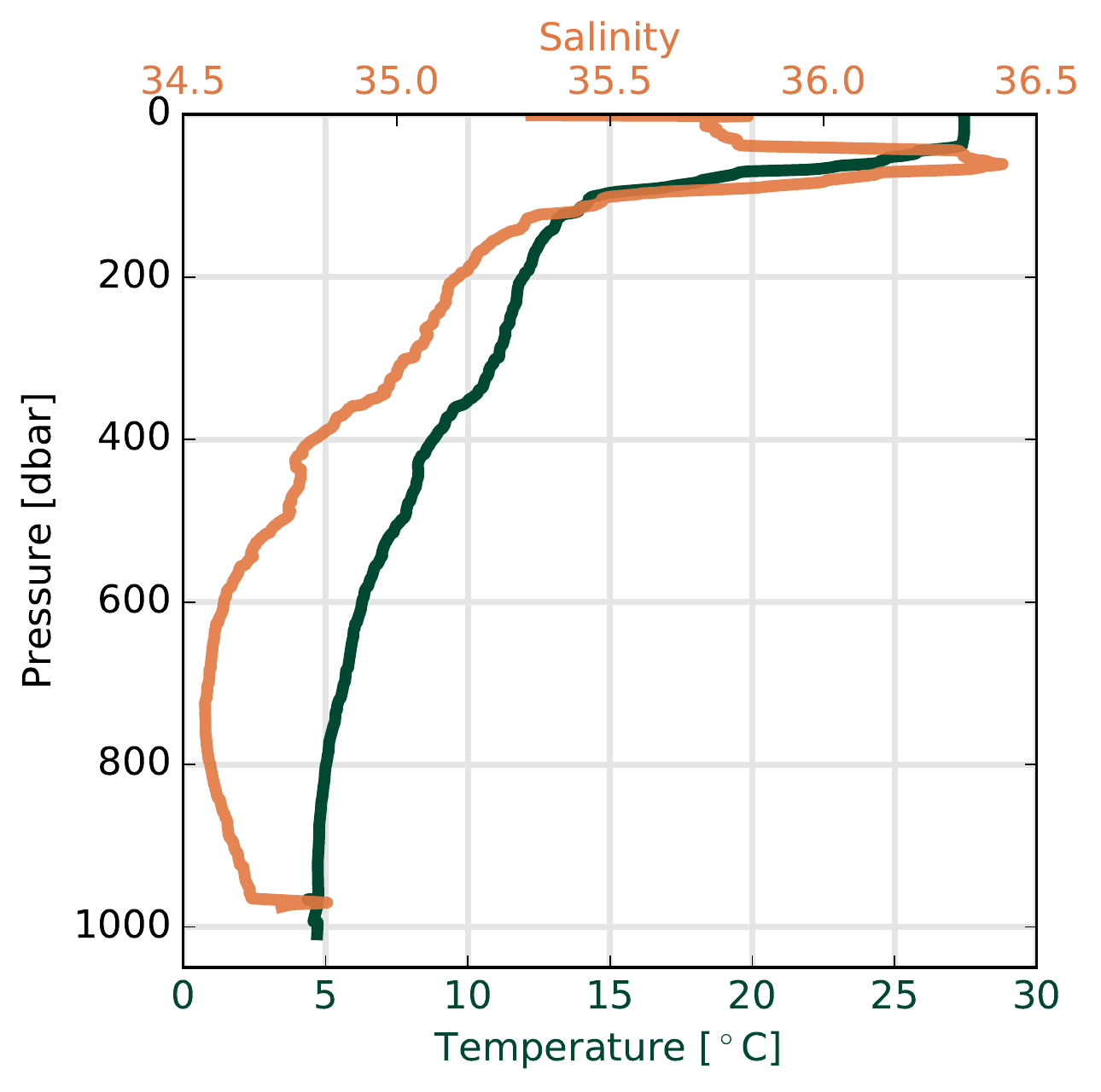}
    \caption{Vertical profiles of temperature (green) and salinity (orange) approximately at 4$^\circ$N 38$^\circ$W on 2008/04/17, from the Brazilian PIRATA hydrographic database. Only the data approved on the quality control procedure recommended by the EuroGOOS is shown. The sharp change of salinity around 950dbar is due to sampling errors, but missed by that quality control. \label{fig:good_TS}}
\end{figure}

The data used to illustrate and discuss CoTeDe is the historical hydrographic CTD\footnote{An electronic sensor that measures conductivity, temperature, and pressure.} dataset from the Prediction and Research Moored Array in the Atlantic (PIRATA) \citep{PIRATA}. 
It is composed of 194 CTD profiles, sampled between 1998 and 2011, with over 380,000 measurements of pressure, temperature, and salinity; 
Figure \ref{fig:good_TS} illustrates one of these profiles. 
The positions of the stations vary along the years, all being nearby the western PIRATA buoys on the western Tropical Atlantic, between 15$^\circ$N 38$^\circ$W
and 19$^\circ$S 34$^\circ$W. 
This dataset is provided by the Brazilian Navy -- Banco Nacional de Dados Oceanogr\'{a}ficos\footnote{https://www.mar.mil.br/dhn/chm/chm\_new/bndo.htm} (BNDO). 

The first task to quality control is to properly extract all the data and metadata available in the CTD output files.
That represents a challenge when using historical data because of the diversity of formats, even from the same manufacturer.
The solution adopted was to create a standalone package, named Seabird\footnote{http:seabird.castelao.net},  to normalize all the data in one common easy--to--use format.
Seabird is an open source Python package, developed with the goal to process the outputs from SBE CTDs and thermosalinographs (TSG).
For each data file, a regular expression that matches up with the content is used to parse the data.
In the case of a new format, the existent regular expressions can be adjusted or a completely new one created, but the common engine is preserved.
With this structure it becomes trivial to extend CoTeDe for a new type of dataset, only requiring a package to parse the raw data on the expected data object. 
Initially developed for CTD, CoTeDe is already extended for TSG and ARGO data.

\subsection{Traditional Quality Control}\label{sec:Traditional}

Oceanographic data have been traditionally quality controlled using a collection of independent tests, each one seeking for a known signature of bad measurements. 
These tests could be grouped in two types: the first one checks for missing, or invalid information, for example, if the measurement is associated with a valid date and valid geographic coordinates; 
The other group of tests compares a characteristic of the data against a previously defined window of acceptable values. 
The most intuitive test on this group is the global range, whose thresholds delimit feasible values in the oceans of the property being considered, for example the temperature itself.
Even though this is a robust criteria, it does not cover spurious data within the range of feasible values. 
One solution to address that is to project the original data onto dimensions that emphasize characteristics of bad measurements, with the goal to obtain a new space where good and bad data spread apart from each other. 
Each projection is hereinafter referred as a \textit{feature} of that measurement. 

On this manuscript I will use $\mathbf{x}$ for a set of measurements of a given variable $x$, that could be temperature for example, and $\mathbf{y_n}$ will be the n$^\mathrm{th}$ feature of $\mathbf{x}$, so that $\mathbf{y_n} = y_n(\mathbf{x})$. 
The index $i$ refers to one specific measurement, like $x_i$, thus $i-1$ ($i+1$) is the previous (following) measurement in the data series. 
Given that, some examples of features widely used are:
\begin{description}
	\item[Rate of Change:] Evaluates the change from the previous measurement as:
	\begin{equation}
		y_r = x_i - x_{i-1}.
            \label{eq:RoC}
	\end{equation}
	\item[Gradient:] Evaluates the rate of change surrounding the measurement, defined as:
        \begin{equation}
		y_g = \left|x_i - \frac{(x_{i+1} + x_{i-1})}{2}\right|.
            \label{eq:Gradient}
        \end{equation}
    \item[Spike:] Evaluates how contrasting a measurement is in comparison with the adjacent successive measurements, defined as:
        \begin{equation}
		y_s = y_g - \left|\frac{(x_{i+1} - x_{i-1})}{2}\right|,
        \end{equation}
	where $y_g$ is the {\it gradient} given by equation \ref{eq:Gradient}.
\item[Tukey 53H:] Evaluates how contrasting a measurement is in comparison to the low frequency tendency of the data series. 
	It takes advantage of the robustness of the median to create a smoother data series which is used as reference \citep{GoringNikora2002}, with the following procedure:
\begin{enumerate}
    \item $x^{(1)}$ is the median of the five points from $x_{i-2}$ to $x_{i+2}$;
    \item $x^{(2)}$ is the median of the three points from $x^{(1)}_{i-1}$ to $x^{(1)}_{i+1}$;
    \item $x^{(3)}$ is defined by the Hanning smoothing filter:
	    \begin{displaymath}
        \frac{1}{4}\left( x^{(2)}_{i-1} +2x^{(2)}_{i} +x^{(2)}_{i+1} \right);
	    \end{displaymath}
    \item Finally:
	    \begin{equation}
		    y_t = \frac{ \left| x_i-x^{(3)} \right|}{\sigma},
	    \end{equation}
\end{enumerate}
where $\sigma$ is the standard deviation of the lowpass filtered data.
    \item[Climatology:] Evaluates the bias between the observed measurement and a climatology, normalized by the expected variability in that point and time, using the relation: 
        \begin{equation}
            y_c = \frac{\left| x_i - \langle x \rangle \right|}{\sigma},
        \end{equation}
	where $\langle x \rangle$ is the climatology, and $\sigma$ is the standard deviation of the observations used to create the climatology.
	Commonly used climatologies are the World Ocean Atlas (WOA) \citep{WOA09Temp, WOA09Sal} and the CSIRO Atlas of Regional Seas (CARS) \citep{Ridgway2002}
\end{description}

\begin{figure}[htb]
    \centering
    \noindent\includegraphics[width=18pc]{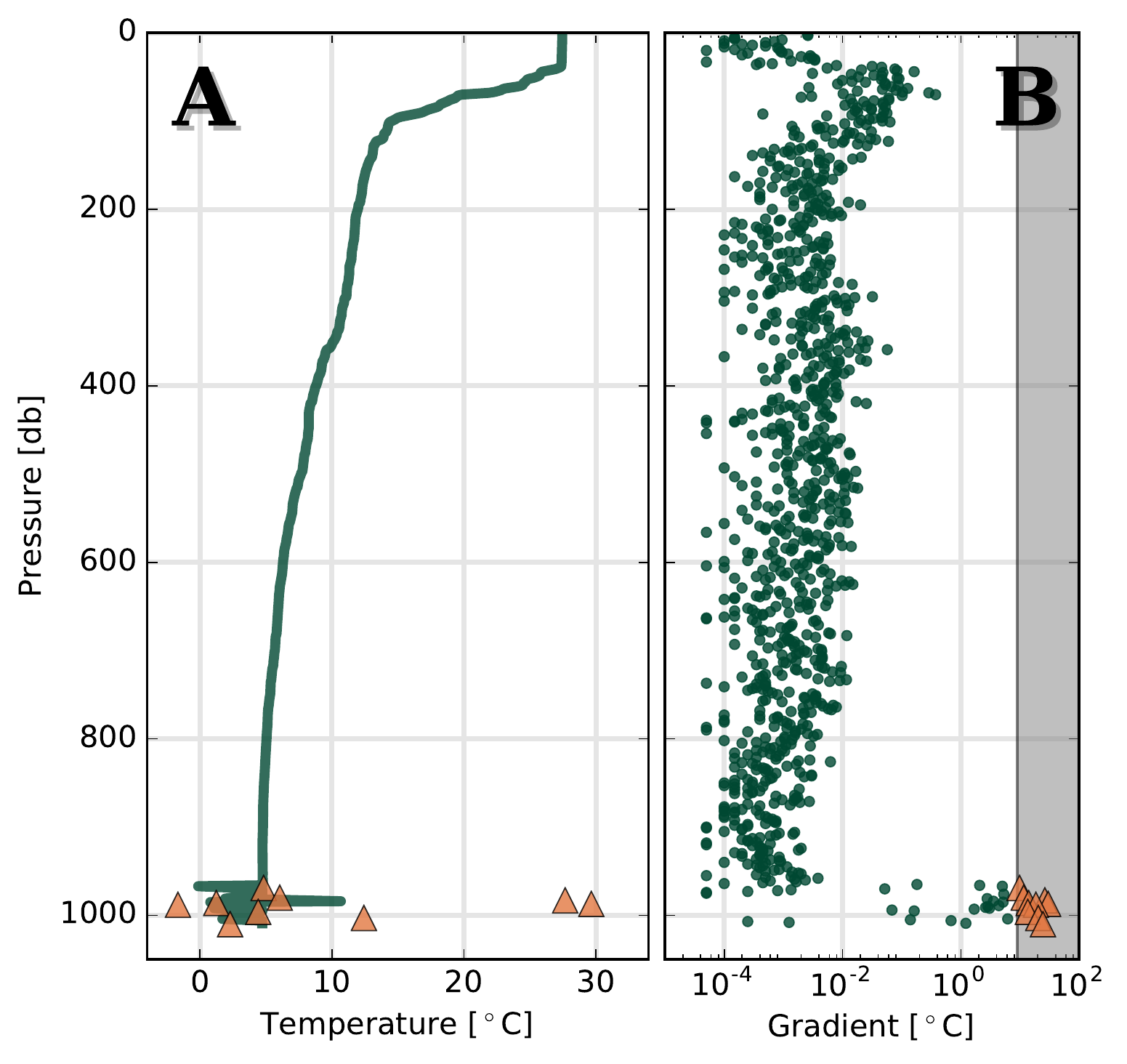}
    \caption{(A) Temperature profile of station \#10 from the PIRATA--X cruise; the green line is the data approved by the gradient test, while the orange triangles failed. (B) The same data plotted in respect to the \textit{gradient} (Eq.: \ref{eq:Gradient}) show a distinct scale for the bad values. The gray area delimits the threshold according to GTSPP.
    \label{fig:gradient_profile}}
\end{figure}

Figure \ref{fig:gradient_profile} illustrates the traditional QC approach with the gradient test, applying a threshold limit to the feature {\it gradient}. 
Data flagged by the global range test was already removed, remaining some spurious measurements near the depth of 1000 dbar. 
The feature {\it gradient} projects the bad data into a distinct scale of the regular temperature observations (see Fig. \ref{fig:gradient_profile}B). 
The variability near 1000 dbar suggests that the threshold used missed some bad data, i.e. some false positive flagging, a subject explored in the following sections.

CoTeDe does not modify or remove any measurement, but returns an overall quality flag for each input value according to the scale recommended by the Intergovernmental Oceanographic Commission of UNESCO (Table \ref{tab:MandatoryFlags}), a widely adopted flag standard \citep{gtsppqc2010, EGOOS2010, Seadatanet10, ARGO2014}.
The final flag of each measurement is the maximum flag value obtained among all performed tests, i.e., it is only considered good (flag 1) if approved by all tests. 
CoTeDe's manual\footnote{http://cotede.castelao.net} provides a full list of implemented tests together with the parameters and thresholds recommended by different groups.

\begin{table}[htb]
	\caption{Quality control flags recommended by IOC--UNESCO, and adopted in CoTeDe. \label{tab:MandatoryFlags}}
	\centering
    \begin{tabular}{c l}
	\hline \hline
	Flag	& Meaning \\
	\hline
        0	& No QC was performed \\
        1	& Good data \\
        2	& Probably good data \\
        3	& Probably bad data \\
        4	& Bad data \\
        6	& Below detection limit \\
        9	& Missing data \\
	\hline \hline
    \end{tabular}
\end{table}

\subsection{Anomaly Detection}\label{sec:AnomalyDetection}

Anomaly detection is a classification technique to discriminate commonly observed data from anomalies. 
While other classification methods try to describe each one of the classes being considered, the anomaly detection approach focusses in recognizing the common data. 
By assuming that spurious measurements are anomalous responses of the sensors, it provides a solution to quality control with less sensitivity to the bad data sample size. 
Further, by avoiding specific patterns to recognize bad data, the anomaly detection promptly identifies unprecedented measuring failures, while other techniques would require to explicitly learn the new pattern first.

This technique is not new to environmental measurements. 
\citet{Bettencourt2007} applied anomaly detection to a network of inland synchronous sensors, identifying bad samples even when they had feasible magnitudes. 
To achieve that, those authors compared each measurement with previous measurements as well as to neighbouring sensors, and detected the anomalies using a p--test.
To evaluate the data by the magnitude itself requires a stationary timeseries or a sampling rate sufficiently high to overcome the environment changing trend \citep{Bettencourt2007}. 
That is an issue for oceanographers, since few marine datasets would meet those requirements, and to aggravate that, 
duplicate measurements in the deep ocean are restricted to modern CTD casts.
From a different argument, \cite{HillMinsker2010} proposed an equivalent concept for the case of individual sensors using sequential measurements into auto--regressive models, including a perceptron type of artificial neural network, to predict the following value in the timeseries.
These authors used a pre--defined limit of confidence on the prediction to obtain the range of tolerance, so a value outside that would be an outlier, and assumed to be a bad measurement. 
This solution can handle small gaps as long as the sampling rate is sufficiently high, but it also requires a stationary timeseries, or regular update on the model parameters. 
Thus, the methodologies used so far in environmental systems are not adequate to quality control oceanographic observations, specially deep ocean measurements.

The alternative that I propose to use anomaly detection in oceanographic data is to project the original variable in dimensions that emphasize different characteristics of the measurement, and then evaluate how anomalous those projections are instead of evaluating the measurement itself. 
The features used in the traditional Q.C. (Section \ref{sec:Traditional}) suits well that task since those were designed to explore known characteristics of bad data. 
Although, instead of testing against fixed thresholds, those are used to characterize the measurement in another scale, for example a {\it gradient} intensity, i.e. the output of the equation \ref{eq:Gradient} (see Fig.: \ref{fig:gradient_profile}B). 
Each feature aggregates a new perspective of the measurement into a multi--dimensional criteria, allowing for a more flexible non--linear classification.
The full procedure implemented in CoTeDe is explained in detail as follows.

\begin{figure}
    \centering
    \noindent\includegraphics[width=18pc]{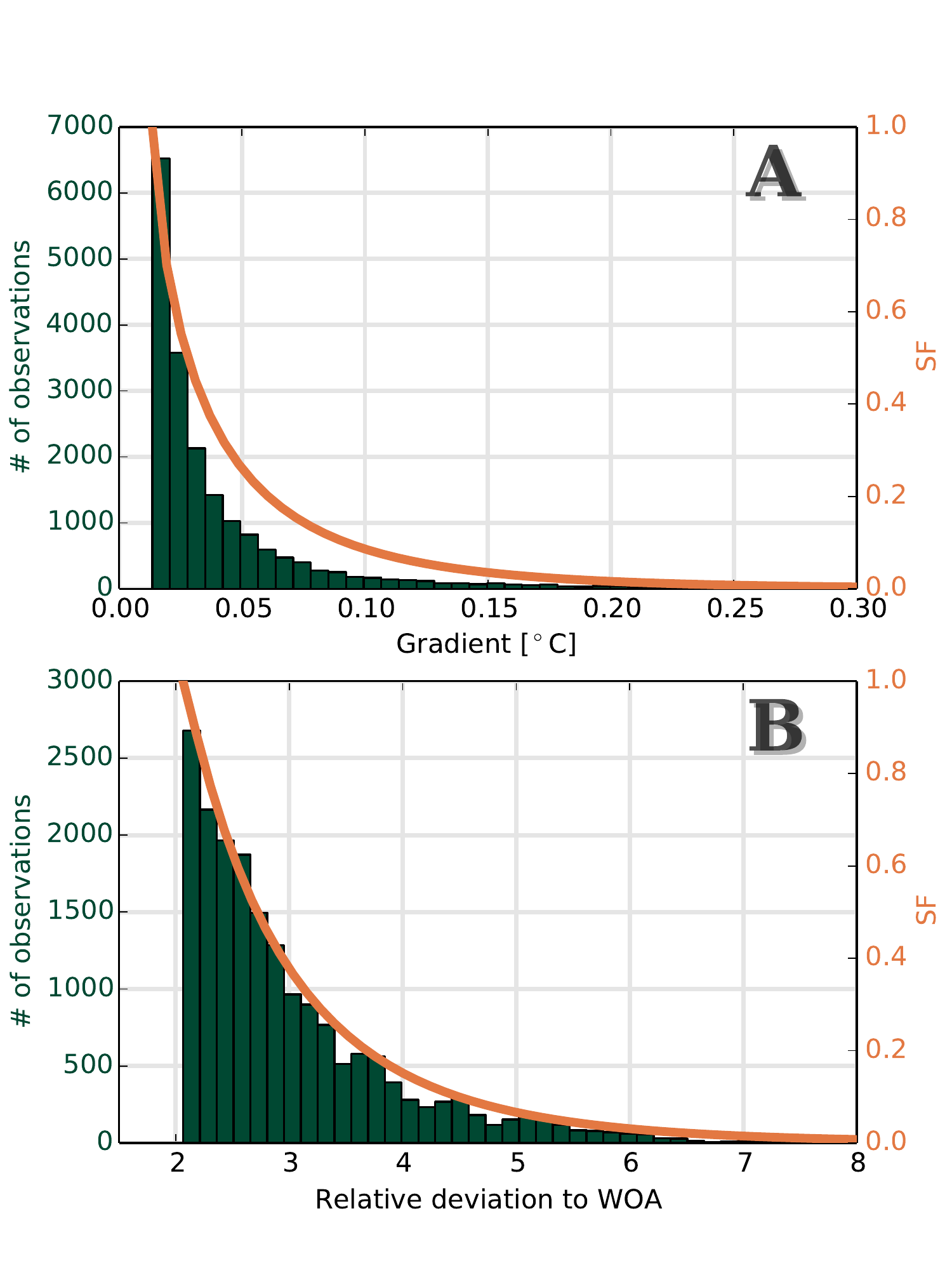}
    \caption{(A) Distribution of the top 10\% {\it gradient} test results (green), and the respective survival function (orange). 
	    (B) Distribution of the top 10\% {\it climatology} comparison test (green), and the respective survival function. 
    Only the data approved by the EuroGOOS QC procedure is considered.\label{fig:SF}}
\end{figure}

The first task is to characterize the typical behavior of the data by estimating a probability density function (PDF) for each feature ($y_n$). 
Since the goal is to identify anomalies, and the bad measurements are usually much less than 1\%, any value below the 90$^\textrm{th}$ percentile is considered common, thus lacking evidence of being a bad measurement. 
The PDF is hence estimated using only the top 10\% values of $\mathbf{y_n}$, allowing a better fit in the range of interest. 
The best results that simultaneously satisfied the different features were obtained from the exponentiated Weibull continuous function, defined as,
\begin{equation}
    \mathrm{PDF}(y|k,\lambda,\alpha) = 
    \alpha \frac{k}{\lambda} 
    \left( \frac{y}{\lambda} \right)^{k-1} \\
    \left[1 - 
        e^{- \left(\frac{y}{\lambda} \right) ^k }
        \right]^{\alpha - 1} 
    e^{- \left(\frac{y}{\lambda} \right) ^k },
\end{equation}
where $k$, $\lambda$, and $\alpha$ are the adjustment parameters, and $y$ is a feature of the variable to be classified, for example the {\it gradient} (Eq: \ref{eq:Gradient}) of measured temperatures: $\mathbf{y_g} = y_g(\mathbf{T})$. 
The respective survival function (SF) of the estimated PDF is used to quantify how anomalous a certain measurement is. 
For a feature $y$, SF$(\mathbf{y}_{i}$) gives approximately how frequent a valid measurement $x$ was observed with $y\ge \mathbf{y}_{i}$. 
Hence the higher $\mathbf{y}_{i}$, the smaller the SF$(\mathbf{y}_{i}$), and more anomalous is $\mathbf{x}_{i}$ in the perspective of the feature $y$. 
Figure \ref{fig:SF} illustrates the top 10\% of {\it gradient} and {\it climatology} of temperature from the PIRATA dataset. 
Only 10\% of the observations had a {\it gradient} over 0.013, therefore, values equal or below that suggest a regular good sample, i.e. such {\it gradient} lacks any indicative of being a bad data. 
In another case, SF$_{\mathrm{g}}(y_g=0.1) = 0.077$, hence there is a 7.7\% chance of obtaining a valid sample $x$ with $y_g \ge 0.1$.

Assuming independent features, the probability of observing a good measurement ($\mathbf{x}_i$), characterized by a set of features $\{y_g(\mathbf{x}_i), y_c(\mathbf{x}_i), y_s(\mathbf{x}_i)\}$, 
or a more rare scenario, is given by the product of the individual probabilities, i.e.
\begin{equation}
	p = \prod_n \mathrm{SF}_n(y_{n}),
	\label{eq:AD_p}
\end{equation}
where $y_n$ is the n$^{\mathrm{th}}$ feature of the measurement $x_i$.

Finally, it is necessary to define a probability threshold ($\tilde p$) in order to distinguish an expected good data from an anomaly. 
To obtain that, the procedure recommended by the EuroGOOS was taken as a good first guess. 
The data approved by the EuroGOOS procedure was randomly split in 3 subgroups: fit, test, and error estimate groups, with 60\%, 20\%, and 20\% of the valid observations respectively. 
The non--approved data was randomly split in half, with each half included in the test and error estimate groups. 
The PDF coefficients were adjusted based only on the fit group, hence, expected to be mostly, if not fully, composed of actual good data, because it is expected a tiny fraction of false positives approved by the EuroGOOS procedure. 
Therefore, the  survival functions are indicative of how common that result is observed among the good data. 
The threshold $\tilde p$ was defined to minimize the sum of false positive and false negative cases considered in the test group. 
The error of the anomaly detection approach was estimated by applying the $\tilde p$ from the previous step on the last data subgroup, the error estimate group.
Since the data on the error estimate group is not used on the adjusting procedures, this is an unbiased error estimate.
In summary, if the probability $\mathbf{p}_i$ is greater than the threshold $\tilde p$, $\mathbf{x}_i$ is flagged as good (1), otherwise it is flagged as probably bad (3).

\subsection{Fuzzy Logic}\label{sec:FuzzyLogic}

In contrast to the typical crisp threshold tests, which results in a binary quality evaluation, the fuzzy logic approach seeks a continuous quality scale, with a fuzzy transition between good and bad data. 
Each measurement is evaluated in a higher dimensional space by combining multiple features together, which allows a classification criterion with more degrees of freedom and a decision with better context awareness. 
One way to fine tune this technique is by adjusting the ranges associated with lower or higher uncertainty. 
That is an outstanding advantage \citep{Morello2014} since the meaning of the adjusting parameters is not hidden in the math of the procedure, like in other techniques, thus the human expert can intuitively associate those parameters with the real world.

A sequel of manuscripts \citep{Timms2011, Morello2011, Morello2014} proposed a fuzzy logic procedure to quality control hydrographic data summarized in the following steps:

\begin{enumerate}
	\item Each variable is evaluated by multiple features of the measurement. 
		\citet{Morello2014} evaluated the water temperature using: {\it climatology}, {\it spike}, and {\it rate of change}\footnote{The {\it rate of change} as defined by \citet{Morello2014} is actually equivalent to the widely used {\it gradient} (eq. \ref{eq:Gradient}), while \citet{Timms2011} defines {\it rate of change} as presented in eq. \ref{eq:RoC}.}, but more features can be added or exchanged keeping the same general procedure. 

	\item Each feature is mapped into three fuzzy sets, i.e. three scales of uncertainty: low, medium, and high. 
		The scaled feature is called membership of the fuzzy set. 
		For example, a temperature measurement identical to the climatology suggests high confidence, therefore, the feature {\it climatology} shall result in a membership equal to 1.0 for low uncertainty and memberships equal to 0.0 for medium and high uncertainty.
		Thus, each measurement results in three memberships times the number of features evaluated.

\item Fuzzy rules group the different fuzzy set into combined memberships for each measurement. 
The high level is combined as the maximum value among all memberships for high uncertainty, while the low and medium levels are each one combined by the mean of its respective memberships. 
While several factors are taken into account to consider some data as good, just one kind of error is sufficient to characterize a bad data, hence, it is the maximum value that leads the decision for the high level of uncertainty. 
At this stage, each measurement is associated to 3 different levels of uncertainty, i.e. 3 values, independent of how many features were evaluated.

\item The traditional flag scale (Table \ref{tab:MandatoryFlags}) is obtained according to:
\begin{itemize}
    \item Flagged as good (1) if the low uncertainty level is higher than 0.9;
    \item Flagged as probably good (2) if low uncertainty level is higher than 0.5 and high uncertainty level is lower than 0.3;
    \item Flagged bad (4) data if a threshold is crossed;
    \item Everything else is flagged potentially correctable (3);
\end{itemize}

\end{enumerate}

Such procedure does not use the medium uncertainty level to obtain the traditional flag scale, so it is actually based in only two levels of uncertainty, low and high. 
The bad data (flag 4) is identified like the traditional QC, therefore, the effective improvement of this implementation is to aggregate this continuous transition through probably good (flag 2) and probably bad (flag 3) data giving more freedom to minimize false positives and false negatives. 

CoTede provides the above--mentioned procedure, along with an alternative implementation of fuzzy logic that effectively uses the medium uncertainty set, allowing for a better resolution in the quality scale. 
Also, it defuzzifies by using the centroid of the combined memberships instead of the step 4 described above. 
The final product is a quality level between 0 and 1 for each measurement. 
To better illustrate those procedures, the Supplementing Material contains some study cases, and for more details on the technique the reader is referred to CoTeDe's manual together with the original manuscripts \citep{Timms2011, Morello2011, Morello2014}.

\section{Results and Discussion}

The quality control procedure for hydrographic data traditionally consists of a sequence of independent tests. 
Although there are different recommendations on which batch of tests to use, the general form is the same: each test checks a feature against a threshold for acceptable values. 
The outcome is hence highly sensitive to the threshold chosen, as a wide (strict) limit favors false positives (negatives). 
Figure \ref{fig:gradient_profile} illustrates that dilemma, where the gradient test misses some spurious measurements near 1000 dbar in order to avoid to misclassify the intense gradient of the thermocline, near the surface.  
Because the gradient test classifies the data 
without any other information than the {\it gradient}, 
the calibration is limited to increase or to reduce the threshold value. 
Since failing in one test is sufficient to flag the data as bad, the strategy usually adopted is to calibrate the tests to minimize false negatives, expecting that the false positives would be identified by another test. 
Therefore, the traditional QC performs well on flagging bad data, and improvements to reduce the burden on human expert QC should target false positives.

The detection of unfeasible values is a trivial procedure, so the real challenge is to identify bad measurements within the range of possible magnitudes. 
Thus, all results and considerations hereinafter are for the PIRATA dataset after discarding the failures on the global range test (Section \ref{sec:Traditional}), 
which removes 0.13\% of the full dataset.

\begin{figure}[htb]
    \centering
    \noindent\includegraphics[width=20pc]{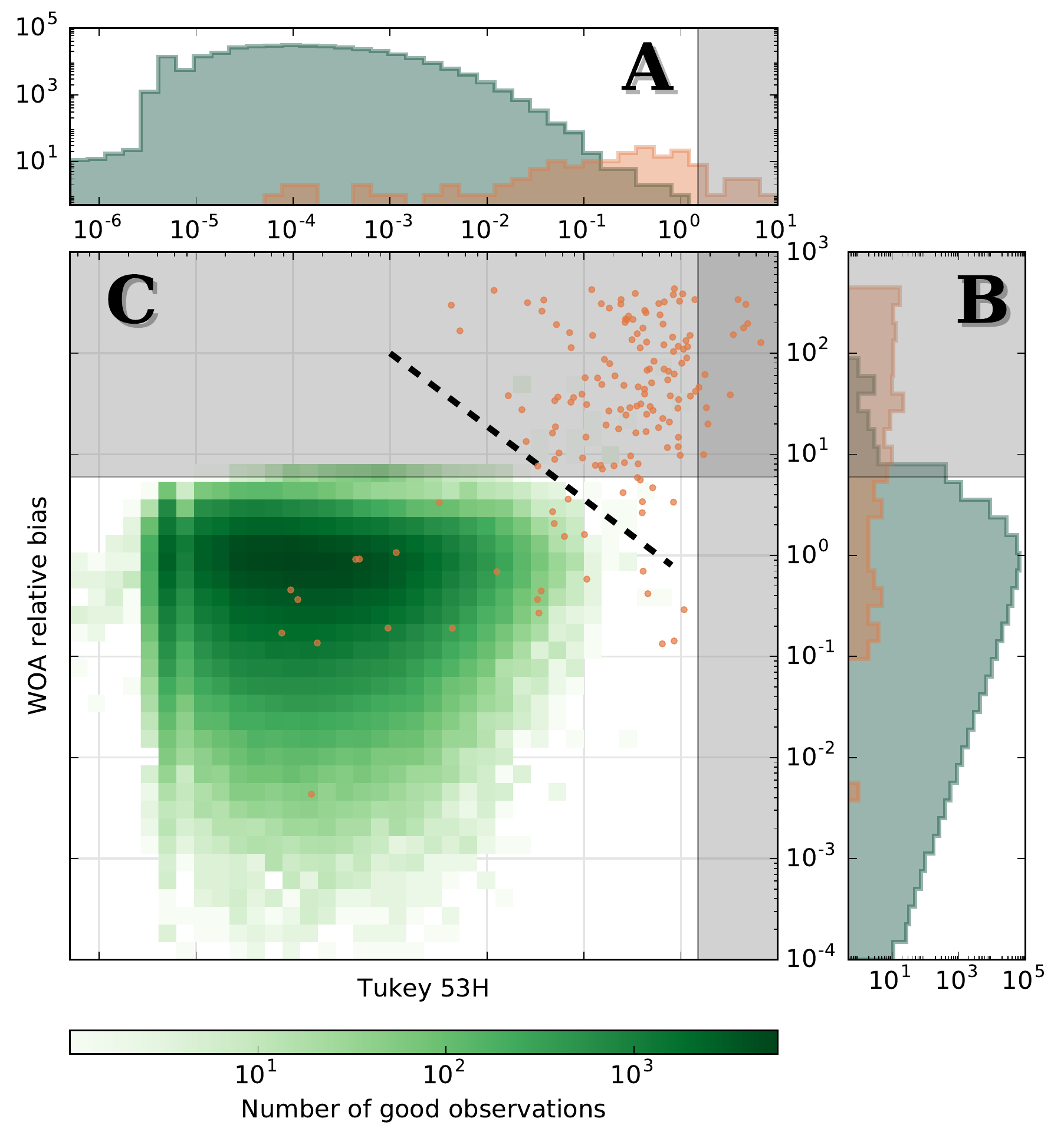}
    \caption{Observations of the PIRATA-Brazil hydrography in respect to {\it Tukey 53H} (A) and {\it climatology} (B). 
	    The good data are in green, and the bad data in orange, according to the EuroGOOS recommendation for realtime. 
	    The gray boxes delimit {\it Tukey 53H} and {\it climatology} above 1.5 and 6, respectively. 
	    The black dashed line is an approximate threshold between the good and bad data clusters.
\label{fig:AnomDet_WOAxTuk_qc}}
\end{figure}

Figure \ref{fig:AnomDet_WOAxTuk_qc} projects the temperature measurements in respect to {\it climatology} and {\it Tukey 53H}. 
The observations are flagged as good (green) or bad (orange) according to EuroGOOS recommendations for realtime data. 
Since that flagging lacks tests with this two features, the classification is independent of the projected dimensions.
The gray rectangles delimit {\it climatology} over 6 and {\it Tukey 53H} over 1.5, illustrating the traditional QC procedure. 
Tukey 53H test agrees with EuroGOOS classification capturing only bad data (gray box on Fig. \ref{fig:AnomDet_WOAxTuk_qc}A), thus it would be redundant if aggregated into EuroGOOS, at least with such threshold. 
In contrast to that, the climatology test flags some data otherwise classified as good (gray box on Fig. \ref{fig:AnomDet_WOAxTuk_qc}B). 
Considering only the data already approved in all other tests from EuroGOOS (green), a climatology test with threshold of 6 would flag another 0.06\% of the data as probably bad (flag 3), while a threshold of 3, as recommended by GTSPP, increases that to 1.05\%. 
That is higher than what would be expected for a normally distributed data.
Manual classification confirms that a threshold of 3 (6) results in more than 1\% (0.05\%) of the dataset as false negatives, i.e. the climatology test recurrently flags uncommon real events as bad data. 
The feature {\it climatology} is not equally distributed around 0 as it would be expected, but with a median of 0.3, hence most of those failures were due to warm anomalies. 
This result suggests that datasets quality controlled by the largely used GTSPP standard would attenuate any long term trend, like in the case of global warming. 
The climatology test, as it is, assumes a normally distributed stationary time series. If that is violated, such test would systematically reject good data, modifying the spectrum of the final product. 
Another potential issue is regions with insufficient historical observations to properly represent the local variability. 
Nonetheless, since the climatology test identifies bad data otherwise missed by other tests,  CoTeDe uses {\it climatology} in the anomaly detection procedure, but with two modifications. 
First, the hard limit threshold is increased to 10 standard deviations instead of 3; Second, the standard error of the climatology is 
discounted from the difference between the climatological mean and the measurement.
Regions with fewer observations have larger standard error, i.e. more uncertainty in the climatology, therefore, the comparison should be more tolerant to differences.

While it is hard to define an optimal threshold for each unidimensional projection, due to the superposition between good and bad data (see A and B on Fig. \ref{fig:AnomDet_WOAxTuk_qc}), the bidimensional space shows a clear polarization between the two classes (see Fig. \ref{fig:AnomDet_WOAxTuk_qc}C). 
The black dashed line is a better criterion than the gray rectangles to classify the data, but such slope is not possible when evaluating the features one at a time. 
The traditional QC procedure is equivalent to widen or to shrink the gray rectangles, but always keeping the same shape. 
The upper edge of the good data cluster (in green on Fig. \ref{fig:AnomDet_WOAxTuk_qc}C) would be flagged as bad by the traditional approach due to {\it climatology} above 6, but manual classification points those as valid measurements from anomalous natural events.
A major advantage of the expert QC comes from the context awareness \citep{Smith2012}, by considering more information to evaluate cases not obvious at first glance. 
Techniques like anomaly detection, fuzzy logic and Bayesian Networks combine features into a multidimensional criterium achieving a superior skill than multiple unidimensional tests. 
The sparse bad data (orange dots, Fig. \ref{fig:AnomDet_WOAxTuk_qc}C) in the middle of the cluster of good data (green cloud, Fig. \ref{fig:AnomDet_WOAxTuk_qc}C), with small values of {\it climatology} and {\it Tukey 53H}, illustrates how the projections can be orthogonal, thus reinforcing the demand on multiple features to identify spurious data. 
A multidimensional space analysis allows for a criterium with more degrees of freedom, and, 
with a careful set of features, the good data is identified as a distinct cluster.

\begin{figure}[htb]
    \centering
    \noindent\includegraphics[width=20pc]{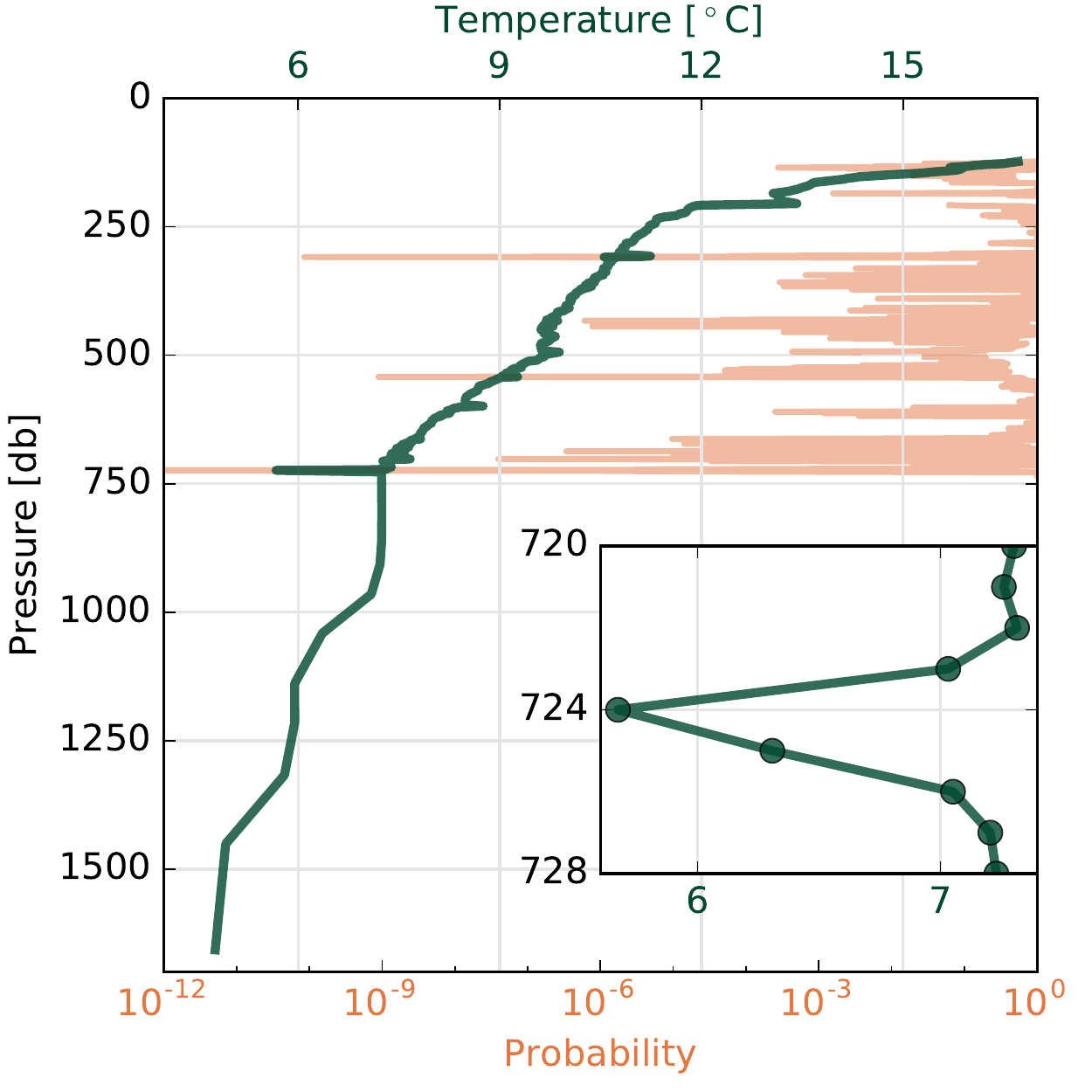}
    \caption{Temperature measurements from the cruise PIRATA--X, profile 10. The data approved by the EuroGOOS procedure is shown in green, and the probability of being a good data, according to the anomaly detection, in orange. The small panel shows a zoom in the temperature between 720 and 728 dbar.\label{fig:TradFailure}}
\end{figure}

Figure \ref{fig:TradFailure} illustrates a profile of temperature approved by
the EuroGOOS criteria. 
The zoom around 724 dbar shows a questionable abrupt change on the profile. 
The features are not large enough to be individually considered bad data by the traditional QC thresholds (see Table \ref{tab:TradFailure}), but the anomaly detection approach identifies this measurement as a distinct structure in the profile (see Figure \ref{fig:TradFailure} in orange). 
While traditional QC does a good job avoiding false negatives, the anomaly detection technique complements that by identifying false positives without requiring a large sample of bad data for training, neither suffering from the imbalance in the dataset.

\begin{table}
\centering
\caption{Observed temperatures and respective quality control test results for the samples 600 to 602 of the profile 10 of cruise PIRATA--X. This interval is also shown in the zoom of Figure \ref{fig:TradFailure}. The last column shows the thresholds suggested by the EuroGOOS procedures. \label{tab:TradFailure}}
\begin{tabular}{l r r r | c}
	\hline \hline
			& $x_{600}$	& $x_{601}$	& $x_{602}$	& thr.\\
	\hline
	Pressure [dbar]	& 723	& 724	& 725	\\
	Temp. [$^\circ$C]	& 7.03 & 5.67	& 6.31	\\
	\hline
	Gradient	& 0.54	& 1.00	& 0.05	& 3$^\circ$ \\
	Spike		& -0.28	& 0.64	& -0.64	& 2$^\circ$ \\
	Climatology	& 0.75	& 3.11	& 1.28	& 6 \\
	Tukey 53H	& 0.01	& 0.28	& 0.15	& 1.5 \\
	\hline
	Anom. det.	& 2e-6	& 1e-20	& 9e-7 \\
	\hline \hline
\end{tabular}
\end{table}

The full EuroGOOS procedure, which includes the climatology test, can be reproduced by the anomaly detection, when calibrated for that purpose, with a mistake rate of approximately 0.4\%. 
Most of the disagreements are due to a known bias for false positives of the traditional QC techniques, as discussed earlier, therefore, that is an overestimate of errors from the anomaly detection. 
A better reference is necessary to properly evaluate the performance of each QC approach, which was obtained through active learning, as follows: 
In a first iteration the anomaly detection was calibrated and evaluated assuming the EuroGOOS as the truth. 
The severity of each supposedly misclassified measurement is quantified by the difference between the probability threshold ($\tilde p$) and the estimated probability of being a good data ($p$). 
The rationale behind it is that to avoid a misclassification with large $|p - \tilde p|$ requires greater changes in the classification parameters that ultimately defined $p$, hence that mistake would be in greater disagreement with the criteria used than one with a small $|p - \tilde p|$. 
That rank of misclassification drives an iterative process where the worst mistake is manually evaluated first and the flag is confirmed or corrected, so the reference is updated, the anomaly detection recalibrated, and the misclassification rank redefined. 
The human effort is hence optimized into classifying first the most critical errors, while the calibration converges. 
The performance of each recalibration is evaluated using an independent dataset, the error subset described in Section \ref{sec:AnomalyDetection}, and the iteration process ceases once the error on the error subset stabilizes or increases, hence, avoiding an over--fitting.
Such active learning results in a better reference classification without manually processing the whole dataset.

Table \ref{tab:QCperformance} shows the performance of the different QC procedures available in CoTeDe. 
The GTSPP and EuroGOOS for realtime (without climatology) achieved the lowest rate of false negatives, but also had the highest rate of false positives. 
To aggregate the climatology on those procedures reduces the ratio of false positives, while increasing the rate of false negatives. 
For the GTSPP that was critical, resulting in the worst overall performance, misclassify 1\% of the dataset.
The two fuzzy logic approaches had a surprising high rate of false negatives, which could probably be improved by a better calibration schema. 
The anomaly detection achieved the best performance overall, with 2 misclassifications per 10,000 measurements, thus, reducing by half the errors by EuroGOOS for realtime, the former best procedure.
It is worth noting that the error by the anomaly detection was estimated from the error subset, hence independent of the data used on the calibration.

\begin{table*}[tb]
    \centering	
    \caption{Ratio of errors per 10,000 measurements of different QC approaches estimated on the PIRATA hydrography dataset approved on the global range test, i.e. after removing the unfeasible values. For reference, manual QC resulted the ratio of 8.8 bad data per 10,000 measurements. \label{tab:QCperformance}}
    \begin{tabular}{l r r r}
	    \hline \hline
	    QC procedure	& False bad	& False good	& Total error \\
	    \hline
	    GTSPP (w/ clim.)	& 97.4		& 2.9		& 100.3 \\
	    GTSPP realtime	& 0.0 		& 5.2		& 5.2 \\
	    EuroGOOS (w/ clim.)	& 4.3 		& 2.8 		& 7.1 \\
	    EuroGOOS realtime	& 0.0 		& 4.1 		& 4.1 \\
	    Morello 2014	& 11.6 		& 2.7 		& 14.3 \\
	    Fuzzy Logic		& 12.0 		& 2.5 		& 14.5 \\
	    Anomaly detection	& 0.2		& 1.8 		& 2.0 \\
	    \hline \hline
    \end{tabular}
\end{table*}

\citet{Gronell2008} also explored the idea of identifying bad data by searching for outliers in multiple features, using an equivalent of the climatology test from the traditional QC, but applied on each feature instead of on the measurement itself. 
That was a major improvement over the traditional QC 
since the local variability scaled the test threshold, avoiding to use one constant for the whole ocean, known to be heterogeneous. 
Despite the differences in the methodologies, it is easy to see some conceptual equivalence with the technique that I propose. 
Some improvements from the anomaly detection do not assume a normal distribution and drastically reduce the manual QC effort, but the effective main advantage comes from the multidimensional criterium.

The Q.C. methodology introduced here allows to include new features, so each feature aggregates a new perspective of the data that can help to identify sampling errors.
For example, some specific cruises analyzed here had a persistent lack of data on the first tens of meters, near the surface, as well as a lowering speed faster than the recommended for CTDs. 
A human expert would note the improper operating procedures, being more likely to flag data as bad on the smallest indication. 
The anomaly detection could mimic that analysis by aggregating two new features: shallowest measurement in a profile, and descending rate. 
\citet{Morello2014} uses for other purposes the time since the last calibration, while \citet{Gronell2008} introduce several other features, which could all be added in this example. 
In case any of those characteristics are too off the expected, that would contribute for the total uncertainty probability ($p$, Eq. \ref{eq:AD_p}), hence a smaller {\it gradient} or {\it spike} would be sufficient to exceed the acceptable $\tilde p$.
The anomaly detection as proposed here is a quantitative way to accumulate different aspects of the data for a non--linear classification, thus making decisions with deeper context awareness.

An intrinsic byproduct of the anomaly detection approach is to define how uncommon is a given scenario. 
\citet{YaoEtal2010} discuss the importance of identifying realtime anomalous, but valid, measurements for management response, like to detect an algae bloom. 
This concept raises new possibilities for autonomous sampling systems.
An intelligent sensor running an onboard realtime quality control could be setup to increase the sampling rate once a threshold on the probability of occurrence is reached.
That would minimize the losses from bad samples, as well as increase the sampling resolution of uncommon events.
It would be a major improvement on the optimization effort of the observing systems.
For example, an underwater glider could stop in a place for one or two dives, before keeping its pre--planned mission, once it detected something different.
An ARGO float could anticipate its cycle and redo a profile if the previous measurements were unexpected.
It is common to keep subsurface moorings over a year at sea without any communication, so any interesting event would only be found after recovering the equipment.
An intelligent adjusting sampling ratio would increase the spectrum coverage of autonomous sensors, with the same storage memory and power budget.

\section{Concluding Remarks}

Machine learning techniques provide fascinating approaches to automatically classify data by employing reinforced learning, which is based in the principle of training the classification system with some known answers.
Such calibration usually requires sufficient data to statistically represent each class, but the better the measuring procedure, the smaller the amount of bad data.
The oriented undersampling can circumvent the contrast in relative sampling sizes \citep{RahmanSmithTimms2014}, but too few bad data is a serious limitation for the usual approach of identifying each class, i.e. recognizing a bad measurement in the same way that it recognizes a good measurement. 
To aggravate that, observations in the open ocean are typically sparse in space and time, allowing for few, if any, overlapping measurements. 
Thus, most machine learning techniques, such as Bayesian Networks and Support Vector Machines, although powerful, are not the most adequate approach for open water oceanographic data due to the intrinsic characteristics of such dataset.

The novel approach based on the anomaly detection technique that I propose strongly impacts the QC of oceanographic data in twofold. 
First it optimizes the expert effort by driving the manual evaluation into the most dubious measurements first, allowing the experts to efficiently handle the increasing amount of measurements in the oceans. 
Second, it combines multiple characteristics of each measurement for a deeper decision making, resulting in a higher context awareness for more intricate classification.
The same Anomaly Detection is not limited to QC, but could also be used to guide self adjusting sampling platforms, increasing the spectrum coverage of the measurements.

The Python package CoTeDe is an open source platform to allow easy application of the current state of the art QC techniques on oceanographic measurements, with the possibility to customize the set of tests to be used.

\section*{Software Availability}
\noindent
\begin{tabular}{l l}
        Package name:           & CoTeDe \\
        Program language:       & Python \\
        Developer:              & Guilherme P. Castel\~{a}o \\
        Available since:        & 2013 \\
        Access:                 & Open source \\
        Website:                & http://cotede.castelao.net \\
        Cost:                   & Free software \\
        License:                & 3--clause BSD\\
\end{tabular}

\section*{Acknowledgments}

CoTeDe was developed while I was funded by the S\~{a}o Paulo Research Foundation (FAPESP) on grant 2013/11825-0.
Luiz Irber and Andrew Ng had great influence on improving CoTeDe.
The graphs on this manuscript and the human training system were built using \citet{Hunter2007}.
Thanks to Bia Villas Boas for the suggestions on this manuscript.
The full name of this package is CoTe De l'eau,
a french name as a tribute to my pleasant extended visit to LEGOS/Toulouse in 2013.

\bibliographystyle{model2-names}
\bibliography{Castelao_cotede}

\begin{thebibliography}{33}
\expandafter\ifx\csname natexlab\endcsname\relax\def\natexlab#1{#1}\fi
\providecommand{\url}[1]{\texttt{#1}}
\providecommand{\href}[2]{#2}
\providecommand{\path}[1]{#1}
\providecommand{\DOIprefix}{doi:}
\providecommand{\ArXivprefix}{arXiv:}
\providecommand{\URLprefix}{URL: }
\providecommand{\Pubmedprefix}{pmid:}
\providecommand{\doi}[1]{\href{http://dx.doi.org/#1}{\path{#1}}}
\providecommand{\Pubmed}[1]{\href{pmid:#1}{\path{#1}}}
\providecommand{\bibinfo}[2]{#2}
\ifx\xfnm\relax \def\xfnm[#1]{\unskip,\space#1}\fi
\bibitem[{Antonov et~al.(2010)Antonov, Seidov, Boyer, Locarnini, Mishonov,
  Garcia, Baranova, Zweng and Johnson}]{WOA09Sal}
\bibinfo{author}{Antonov, J.I.}, \bibinfo{author}{Seidov, D.},
  \bibinfo{author}{Boyer, T.P.}, \bibinfo{author}{Locarnini, R.A.},
  \bibinfo{author}{Mishonov, A.V.}, \bibinfo{author}{Garcia, H.E.},
  \bibinfo{author}{Baranova, O.K.}, \bibinfo{author}{Zweng, M.M.},
  \bibinfo{author}{Johnson, D.R.}, \bibinfo{year}{2010}.
\newblock \bibinfo{title}{Salinity}, in:  \cite{WOA09}.
\newblock \bibinfo{note}{184 pp.}
\bibitem[{{Backer Jr.}(1981)}]{Baker1981}
\bibinfo{author}{{Backer Jr.}, J.}, \bibinfo{year}{1981}.
\newblock \bibinfo{title}{Ocean instruments and experiment design}, in:
  \bibinfo{editor}{Warren, B.A.}, \bibinfo{editor}{Wunch, C.} (Eds.),
  \bibinfo{booktitle}{Evolution of physical oceanography}.
  \bibinfo{publisher}{MIT press}. chapter~\bibinfo{chapter}{I4}, pp.
  \bibinfo{pages}{396--433}.
\bibitem[{Bailey et~al.(1994)Bailey, Gronell, Phillips, Tanner and
  Meyers}]{Bailey1994}
\bibinfo{author}{Bailey, R.}, \bibinfo{author}{Gronell, A.},
  \bibinfo{author}{Phillips, H.}, \bibinfo{author}{Tanner, E.},
  \bibinfo{author}{Meyers, G.}, \bibinfo{year}{1994}.
\newblock \bibinfo{title}{Quality control cookbook for xbt data, version 1.1}.
\newblock \bibinfo{journal}{CSIRO Marine Laboratories Reports}
  \bibinfo{volume}{221}.
\bibitem[{Barker et~al.(2011)Barker, Dunn, Domingues and Wijffels}]{Barker2011}
\bibinfo{author}{Barker, P.M.}, \bibinfo{author}{Dunn, J.R.},
  \bibinfo{author}{Domingues, C.M.}, \bibinfo{author}{Wijffels, S.E.},
  \bibinfo{year}{2011}.
\newblock \bibinfo{title}{Pressure sensor drifts in argo and their impacts}.
\newblock \bibinfo{journal}{Journal of Atmospheric and Oceanic Technology}
  \bibinfo{volume}{28}, \bibinfo{pages}{1036--1049}.
\bibitem[{Bettencourt et~al.(2007)Bettencourt, Hagberg and
  Larkey}]{Bettencourt2007}
\bibinfo{author}{Bettencourt, L.}, \bibinfo{author}{Hagberg, A.},
  \bibinfo{author}{Larkey, L.}, \bibinfo{year}{2007}.
\newblock \bibinfo{title}{{Separating the Wheat from the Chaff: Practical
  Anomaly Detection Schemes in Ecological Applications of Distributed Sensor
  Networks}}.
\newblock \bibinfo{journal}{Distrib. Comput. Sens. Syst.}
  \bibinfo{volume}{4549}, \bibinfo{pages}{223--239}.
\newblock \DOIprefix\doi{10.1007/978-3-540-73090-3\_15}.
\bibitem[{Cowley et~al.(2013)Cowley, Wijffels, Cheng, Boyer and
  Kizu}]{Cowley2013}
\bibinfo{author}{Cowley, R.}, \bibinfo{author}{Wijffels, S.},
  \bibinfo{author}{Cheng, L.}, \bibinfo{author}{Boyer, T.},
  \bibinfo{author}{Kizu, S.}, \bibinfo{year}{2013}.
\newblock \bibinfo{title}{Biases in expendable bathythermograph data: A new
  view based on historical side-by-side comparisons}.
\newblock \bibinfo{journal}{Journal of Atmospheric and Oceanic Technology}
  \bibinfo{volume}{30}, \bibinfo{pages}{1195--1225}.
\bibitem[{{DATA--MEQ working group}(2010)}]{EGOOS2010}
\bibinfo{author}{{DATA--MEQ working group}}, \bibinfo{year}{2010}.
\newblock \bibinfo{title}{Recommendations for \emph{in--situ} data Near Real
  Time Quality Control}. \bibinfo{edition}{eg10.19} ed.
\newblock \bibinfo{organization}{European Global Ocean Observing System}.
\newblock \bibinfo{note}{EG10.19}.
\bibitem[{Domingues et~al.(2008)Domingues, Church, White, Gleckler, Wijffels,
  Barker and Dunn}]{Domingues2008}
\bibinfo{author}{Domingues, C.M.}, \bibinfo{author}{Church, J.A.},
  \bibinfo{author}{White, N.J.}, \bibinfo{author}{Gleckler, P.J.},
  \bibinfo{author}{Wijffels, S.E.}, \bibinfo{author}{Barker, P.M.},
  \bibinfo{author}{Dunn, J.R.}, \bibinfo{year}{2008}.
\newblock \bibinfo{title}{Improved estimates of upper-ocean warming and
  multi--decadal sea--level rise}.
\newblock \bibinfo{journal}{Nature} \bibinfo{volume}{453},
  \bibinfo{pages}{1090--1093}.
\bibitem[{Feistel(2008)}]{Feistel2008}
\bibinfo{author}{Feistel, R.}, \bibinfo{year}{2008}.
\newblock \bibinfo{title}{A gibbs function for seawater thermodynamics for- 6
  to 80° c and salinity up to 120gkg--1}.
\newblock \bibinfo{journal}{Deep Sea Research Part I: Oceanographic Research
  Papers} \bibinfo{volume}{55}, \bibinfo{pages}{1639--1671}.
\bibitem[{Gill(1982)}]{Gill1982}
\bibinfo{author}{Gill, A.E.}, \bibinfo{year}{1982}.
\newblock \bibinfo{title}{{Atmosphere-Ocean Dynamics}}.
\newblock \bibinfo{edition}{1} ed., \bibinfo{publisher}{Academic Press}.
\bibitem[{Goring and Nikora(2002)}]{GoringNikora2002}
\bibinfo{author}{Goring, D.G.}, \bibinfo{author}{Nikora, V.I.},
  \bibinfo{year}{2002}.
\newblock \bibinfo{title}{Despiking acoustic doppler velocimeter data}.
\newblock \bibinfo{journal}{Journal of Hydraulic Engineering}
  \bibinfo{volume}{128}, \bibinfo{pages}{117--126}.
\bibitem[{Gronell and Wijffels(2008)}]{Gronell2008}
\bibinfo{author}{Gronell, A.}, \bibinfo{author}{Wijffels, S.E.},
  \bibinfo{year}{2008}.
\newblock \bibinfo{title}{A semiautomated approach for quality controlling
  large historical ocean temperature archives}.
\newblock \bibinfo{journal}{J. Atmos. Ocean. Technol.} \bibinfo{volume}{25},
  \bibinfo{pages}{990--1003}.
\newblock \DOIprefix\doi{10.1175/JTECHO539.1}.
\bibitem[{Hill and Minsker(2010)}]{HillMinsker2010}
\bibinfo{author}{Hill, D.J.}, \bibinfo{author}{Minsker, B.S.},
  \bibinfo{year}{2010}.
\newblock \bibinfo{title}{Anomaly detection in streaming environmental sensor
  data: A data-driven modeling approach}.
\newblock \bibinfo{journal}{Environmental Modelling \& Software}
  \bibinfo{volume}{25}, \bibinfo{pages}{1014--1022}.
\bibitem[{Hunter(2007)}]{Hunter2007}
\bibinfo{author}{Hunter, J.D.}, \bibinfo{year}{2007}.
\newblock \bibinfo{title}{Matplotlib: A 2d graphics environment}.
\newblock \bibinfo{journal}{Computing In Science \& Engineering}
  \bibinfo{volume}{9}, \bibinfo{pages}{90--95}.
\bibitem[{Ivezi{\'c} et~al.(2014)Ivezi{\'c}, Connolly, VanderPlas and
  Gray}]{IvezicConnollyVanderPlas2014}
\bibinfo{author}{Ivezi{\'c}, {\v{Z}}.}, \bibinfo{author}{Connolly, A.J.},
  \bibinfo{author}{VanderPlas, J.T.}, \bibinfo{author}{Gray, A.},
  \bibinfo{year}{2014}.
\newblock \bibinfo{title}{Statistics, Data Mining, and Machine Learning in
  Astronomy: A Practical Python Guide for the Analysis of Survey Data}.
\newblock \bibinfo{publisher}{Princeton University Press}.
\bibitem[{Levitus(2010)}]{WOA09}
\bibinfo{editor}{Levitus, S.} (Ed.), \bibinfo{year}{2010}.
\newblock \bibinfo{publisher}{NOAA Atlas NESDIS 69, U.S. Government Printing
  Office}, \bibinfo{address}{Washington, D.C.}
\bibitem[{Levitus et~al.(2009)Levitus, Antonov, Boyer, Locarnini, Garcia and
  Mishonov}]{Levitus2009}
\bibinfo{author}{Levitus, S.}, \bibinfo{author}{Antonov, J.},
  \bibinfo{author}{Boyer, T.}, \bibinfo{author}{Locarnini, R.},
  \bibinfo{author}{Garcia, H.}, \bibinfo{author}{Mishonov, A.},
  \bibinfo{year}{2009}.
\newblock \bibinfo{title}{Global ocean heat content 1955--2008 in light of
  recently revealed instrumentation problems}.
\newblock \bibinfo{journal}{Geophysical Research Letters} \bibinfo{volume}{36}.
\bibitem[{Locarnini et~al.(2010)Locarnini, Mishonov, Antonov, Boyer, Garcia,
  Baranova, Zweng and Johnson}]{WOA09Temp}
\bibinfo{author}{Locarnini, R.A.}, \bibinfo{author}{Mishonov, A.V.},
  \bibinfo{author}{Antonov, J.I.}, \bibinfo{author}{Boyer, T.P.},
  \bibinfo{author}{Garcia, H.E.}, \bibinfo{author}{Baranova, O.K.},
  \bibinfo{author}{Zweng, M.M.}, \bibinfo{author}{Johnson, D.R.},
  \bibinfo{year}{2010}.
\newblock \bibinfo{title}{Temperature}, in:  \cite{WOA09}.
\newblock \bibinfo{note}{184 pp.}
\bibitem[{Millero et~al.(2008)Millero, Feistel, Wright and
  McDougall}]{Millero2008}
\bibinfo{author}{Millero, F.J.}, \bibinfo{author}{Feistel, R.},
  \bibinfo{author}{Wright, D.G.}, \bibinfo{author}{McDougall, T.J.},
  \bibinfo{year}{2008}.
\newblock \bibinfo{title}{The composition of standard seawater and the
  definition of the reference-composition salinity scale}.
\newblock \bibinfo{journal}{Deep Sea Research Part I: Oceanographic Research
  Papers} \bibinfo{volume}{55}, \bibinfo{pages}{50--72}.
\bibitem[{Morello et~al.(2011)Morello, Lynch, Slawinski, Howell, Hughes and
  Timms}]{Morello2011}
\bibinfo{author}{Morello, E.}, \bibinfo{author}{Lynch, T.},
  \bibinfo{author}{Slawinski, D.}, \bibinfo{author}{Howell, B.},
  \bibinfo{author}{Hughes, D.}, \bibinfo{author}{Timms, G.},
  \bibinfo{year}{2011}.
\newblock \bibinfo{title}{Quantitative quality control (qc) procedures for the
  australian national reference stations: Sensor data}, in:
  \bibinfo{booktitle}{OCEANS 2011}, \bibinfo{organization}{IEEE},
  \bibinfo{address}{Waikoloa, HI}. pp. \bibinfo{pages}{1--7}.
\bibitem[{Morello et~al.(2014)Morello, Galibert, Smith, Ridgway, Howell,
  Slawinski, Timms, Evans and Lynch}]{Morello2014}
\bibinfo{author}{Morello, E.B.}, \bibinfo{author}{Galibert, G.},
  \bibinfo{author}{Smith, D.}, \bibinfo{author}{Ridgway, K.R.},
  \bibinfo{author}{Howell, B.}, \bibinfo{author}{Slawinski, D.},
  \bibinfo{author}{Timms, G.P.}, \bibinfo{author}{Evans, K.},
  \bibinfo{author}{Lynch, T.P.}, \bibinfo{year}{2014}.
\newblock \bibinfo{title}{{Quality Control (QC) procedures for Australia’s
  National Reference Station’s sensor data—Comparing semi-autonomous
  systems to an expert oceanographer}}.
\newblock \bibinfo{journal}{Methods Oceanogr.} \bibinfo{volume}{9},
  \bibinfo{pages}{17--33}.
\newblock \DOIprefix\doi{10.1016/j.mio.2014.09.001}.
\bibitem[{{QARTOD group}(2016)}]{QARTOD2016}
\bibinfo{author}{{QARTOD group}}, \bibinfo{year}{2016}.
\newblock \bibinfo{title}{Manual for Real--Time Quality Control of In-situ
  Temperature and Salinity Data}. \bibinfo{edition}{version 2.0} ed.
\newblock \bibinfo{organization}{IOOS}.
\bibitem[{Rahman et~al.(2014)Rahman, Member, Smith and
  Timms}]{RahmanSmithTimms2014}
\bibinfo{author}{Rahman, A.}, \bibinfo{author}{Member, S.},
  \bibinfo{author}{Smith, D.V.}, \bibinfo{author}{Timms, G.},
  \bibinfo{year}{2014}.
\newblock \bibinfo{title}{{Quality Assessment of Sensor Data}}.
\newblock \bibinfo{journal}{IEEE Sensors Journal} \bibinfo{volume}{14},
  \bibinfo{pages}{1035--1047}.
\bibitem[{Rahman et~al.(2013)Rahman, Smith and Timms}]{Rahman2013}
\bibinfo{author}{Rahman, A.}, \bibinfo{author}{Smith, D.V.},
  \bibinfo{author}{Timms, G.}, \bibinfo{year}{2013}.
\newblock \bibinfo{title}{Multiple classifier system for automated quality
  assessment of marine sensor data}, in: \bibinfo{booktitle}{Intelligent
  Sensors, Sensor Networks and Information Processing, 2013 IEEE Eighth
  International Conference on}, \bibinfo{organization}{IEEE}. pp.
  \bibinfo{pages}{362--367}.
\bibitem[{Ridgway et~al.(2002)Ridgway, Dunn and Wilkin}]{Ridgway2002}
\bibinfo{author}{Ridgway, K.R.}, \bibinfo{author}{Dunn, J.R.},
  \bibinfo{author}{Wilkin, J.L.}, \bibinfo{year}{2002}.
\newblock \bibinfo{title}{Ocean interpolation by four-dimensional weighted
  least squares--application to the waters around {A}ustralasia}.
\newblock \bibinfo{journal}{Journal of Atmospheric and Oceanic Technology}
  \bibinfo{volume}{19}, \bibinfo{pages}{1357--1375}.
\newblock \DOIprefix\doi{10.1175/1520-0426(2002)0192.0.CO;2}.
\bibitem[{SeaDataNet(2010)}]{Seadatanet10}
\bibinfo{author}{SeaDataNet}, \bibinfo{year}{2010}.
\newblock \bibinfo{title}{Data Quality Control Procedures}.
  \bibinfo{edition}{version 2.0} ed.
\newblock \bibinfo{organization}{SeaDatanet}.
\bibitem[{Servain et~al.(1998)Servain, Busalacchi, McPhaden, Moura
  et~al.}]{PIRATA}
\bibinfo{author}{Servain, J.}, \bibinfo{author}{Busalacchi, A.},
  \bibinfo{author}{McPhaden, M.J.}, \bibinfo{author}{Moura, A.D.}, et~al.,
  \bibinfo{year}{1998}.
\newblock \bibinfo{title}{A pilot research moored array in the tropical
  atlantic ({PIRATA})}.
\newblock \bibinfo{journal}{Bulletin of the American Meteorological Society}
  \bibinfo{volume}{79}, \bibinfo{pages}{2019}.
\bibitem[{Smith et~al.(2012)Smith, Timms, De~Souza and D’Este}]{Smith2012}
\bibinfo{author}{Smith, D.}, \bibinfo{author}{Timms, G.},
  \bibinfo{author}{De~Souza, P.}, \bibinfo{author}{D’Este, C.},
  \bibinfo{year}{2012}.
\newblock \bibinfo{title}{A bayesian framework for the automated online
  assessment of sensor data quality}.
\newblock \bibinfo{journal}{Sensors} \bibinfo{volume}{12},
  \bibinfo{pages}{9476--9501}.
\bibitem[{Timms et~al.(2011)Timms, de~Souza, Reznik and Smith}]{Timms2011}
\bibinfo{author}{Timms, G.P.}, \bibinfo{author}{de~Souza, P.a.},
  \bibinfo{author}{Reznik, L.}, \bibinfo{author}{Smith, D.V.},
  \bibinfo{year}{2011}.
\newblock \bibinfo{title}{{Automated data quality assessment of marine
  sensors}}.
\newblock \bibinfo{journal}{Sensors} \bibinfo{volume}{11},
  \bibinfo{pages}{9589--9602}.
\newblock \DOIprefix\doi{10.3390/s111009589}.
\bibitem[{{UNESCO--IOC}, 2010()}]{gtsppqc2010}
{UNESCO--IOC}, 2010, \bibinfo{year}{2010}.
\newblock \bibinfo{title}{GTSPP Real-Time Quality Control Manual}.
  \bibinfo{edition}{first revised edition} ed.
\newblock \bibinfo{organization}{UNESCO--IOC}. \bibinfo{address}{United Nations
  Educational, Scientific and Cultural Organization 7, Place de Fontenoy,
  75352, {P}aris 07 SP}.
\newblock \bibinfo{note}{IOC/2010/MG/22Rev.}
\bibitem[{Wong et~al.(2014)Wong, Keeley, Carval and {Argo Data Management
  Team}}]{ARGO2014}
\bibinfo{author}{Wong, A.}, \bibinfo{author}{Keeley, R.},
  \bibinfo{author}{Carval, T.}, \bibinfo{author}{{Argo Data Management Team}},
  \bibinfo{year}{2014}.
\newblock \bibinfo{title}{Argo Quality Control Manual}.
\newblock \DOIprefix\doi{http://dx.doi.org/10.13155/33951}.
  \bibinfo{note}{version 2.9.1}.
\bibitem[{Wong et~al.(2015)Wong, Keeley, Carval and {Argo Data Management
  Team}}]{ARGO2015}
\bibinfo{author}{Wong, A.}, \bibinfo{author}{Keeley, R.},
  \bibinfo{author}{Carval, T.}, \bibinfo{author}{{Argo Data Management Team}},
  \bibinfo{year}{2015}.
\newblock \bibinfo{title}{Argo Quality Control Manual For {CTD} and Trajectory
  Data}.
\newblock \DOIprefix\doi{http://dx.doi.org/10.13155/33951}.
  \bibinfo{note}{version 3.0}.
\bibitem[{Yao et~al.(2010)Yao, Sharma, Golubchik and Govindan}]{YaoEtal2010}
\bibinfo{author}{Yao, Y.}, \bibinfo{author}{Sharma, A.},
  \bibinfo{author}{Golubchik, L.}, \bibinfo{author}{Govindan, R.},
  \bibinfo{year}{2010}.
\newblock \bibinfo{title}{Online anomaly detection for sensor systems: A simple
  and efficient approach}.
\newblock \bibinfo{journal}{Performance Evaluation} \bibinfo{volume}{67},
  \bibinfo{pages}{1059--1075}.

\end{thebibliography}

\end{document}